\documentclass[letter,fleqn,usenatbib]{mnras}

\usepackage[T1]{fontenc}
\usepackage{ae,aecompl}
\usepackage[switch,modulo]{lineno}

\usepackage{graphicx}	
\usepackage{amsmath}	
\usepackage{amssymb,gensymb}	

\usepackage[caption = false]{subfig} 

\title{NGC 5626: a massive fast rotator with a twist}

\author[S. Viaene et al.]{
S. Viaene$^{1,2}$\thanks{E-mail: sebastien.viaene@ugent.be}, 
M. Sarzi$^{2}$, 
M. Baes$^{1}$,
I. Puerari$^{3}$
\\
$^{1}$Sterrenkundig Observatorium, Universiteit Gent, Krijgslaan 281, B-9000 Gent, Belgium\\
$^{2}$Centre for Astrophysics Research, University of Hertfordshire, College Lane, Hatfield AL10 9AB, UK \\
$^{3}$Instituto Nacional de Astrof\'{i}sica, Optica y Electr\'{o}nica, Calle Luis Enrique Erro 1, 72840 Santa Mar\'{i}a Tonantzintla, Puebla, Mexico
}

\date{Accepted 2017 November 22. Received 2017 November 5; in original form 2017 September 5.}

\pubyear{2017}

\begin{document}
\label{firstpage}
\pagerange{\pageref{firstpage}--\pageref{lastpage}}
\maketitle

\begin{abstract}
We present a kinematic analysis of the dust-lane elliptical NGC 5626 based on MUSE observations. These data allow to robustly classify this galaxy as a fast rotator and to infer a virial mass of $10^{11.7} M_\odot$, making it one of the most massive fast rotators known. In addition, the depth and extent of the MUSE data reveal a strong kinematic twist in the stellar velocity field (by up to $45$ degrees beyond $1.5R_e$).  A comparison with the ATLAS$^\mathrm{3D}$ sample underlines the rareness of this system, although we show that such a large-scale kinematic twist could have been missed by the ATLAS$^\mathrm{3D}$ data due to the limited spatial sampling of this survey (typically extending to $0.6R_e$ for massive ETGs). MUSE thus has the potential to unveil more examples of this type of galaxies. We discuss the environment and possible formation history of NGC 5626 and finally argue how a merger between the Milky Way and Andromeda could produce a galaxy of the same class as NGC 5626.
\end{abstract}

\begin{keywords}
galaxies: individual: NGC 5626 -- galaxies: fundamental: parameters --
\end{keywords}



\section{Introduction}

The classification of early-type galaxies (ETGs) has been revolutionised in the past decades. The classic E0-E6 classification of \citet{Hubble1936} was largely based on inclination. \citet{Kormendy1996} proposed a photometric classification based on boxiness/disciness of optical isophotes, which better reflects the dynamical state of the galaxy. The advent of integral field spectroscopy surpassed the photometric division of elliptical and lenticular galaxies. Instead, ETGs are now considered to be either fast or slow rotators (see e.g. \citealt{Emsellem2007, Emsellem2011}, and \citealt{Cappellari2016} for reviews). It has become common practice to classify a galaxy's rotational class based on the $\lambda_R$ parameter, originally proposed by \citet{Emsellem2007}. This metric is a proxy for stellar angular momentum per unit mass, with higher values for fast rotators and lower $\lambda_R$ for slow rotators.

Fast rotators tend to have regular and symmetric velocity fields, and have no misalignment between the photometric and kinematic position angle (PA). They are intrinsically flatter and anisotropic, and often exhibit stellar discs \citep{Krajnovic2011,Weijmans2014}. Fast rotators are generally less massive than slow rotators, but are more numerous and make up around $70 \%$ of the total stellar mass in ETGs \citep{Emsellem2011}.

On the other hand, slow rotators include the most massive ETGs and are found to be intrinsically rounder. Their velocity field are more chaotic, and frequently include twists, decoupled cores, etc. More often than not, there is a misalignment between their kinematic and photometric PA \citep[see e.g.][for a more detailed account of the subclasses of slow rotators]{Krajnovic2011}.

\begin{figure*}
	\includegraphics[width=1.0\textwidth]{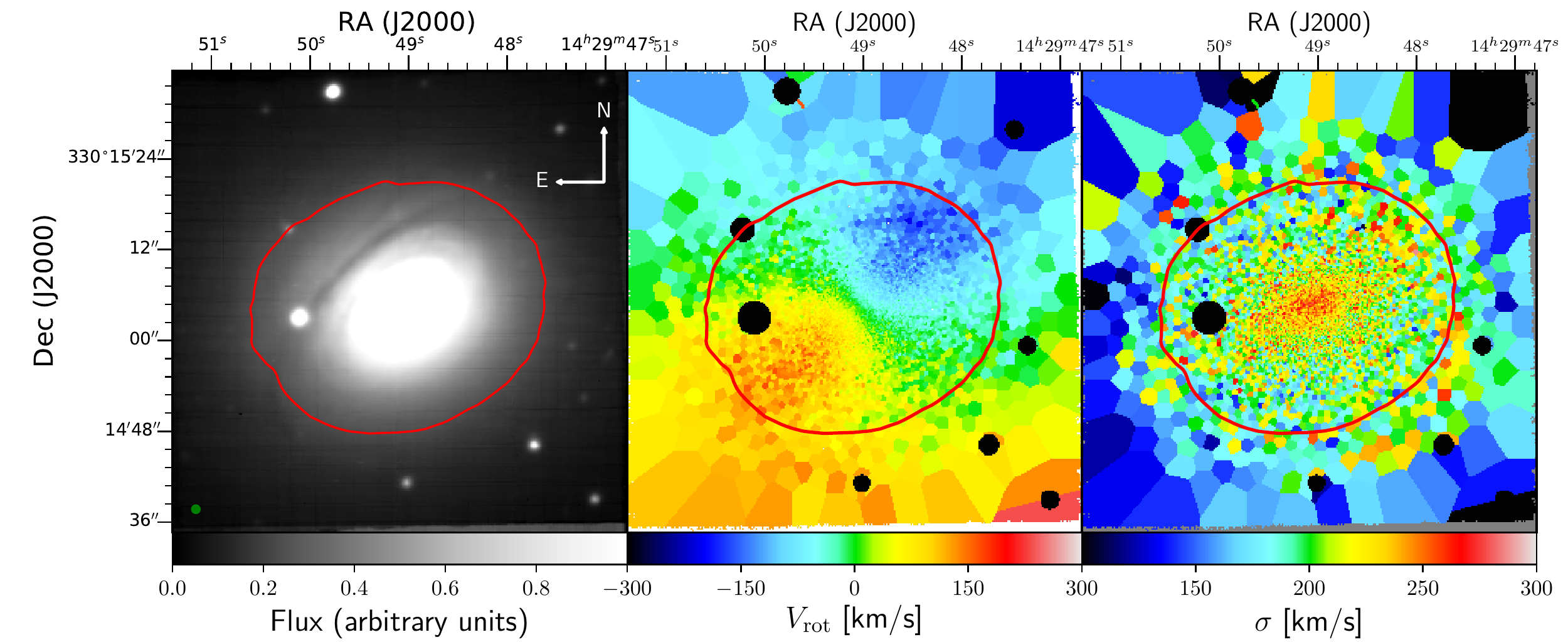}
    \caption{Left: integrated flux of the MUSE IFU spectra of NGC 5626. Middle: Voronoi binned radial velocity field of the stellar component, clearly showing a twist with radius. Right: the corresponding velocity dispersion map. The red contour corresponds to the flux level of one effective radius ($R_e = 18.3$) arcsec. Black circles are masked out foreground stars.}
    \label{fig:maps}
\end{figure*}

The origin of this dichotomy is much debated, and several formation mechanisms have been proposed to create both types of ETGs. The current understanding is that mergers play a crucial role, but the details are unclear. The progenitors of ETGs are likely normal star-forming galaxies. Gas-rich minor mergers maintain the galaxy's angular momentum, while dry minor mergers trigger disc instabilities and transform the galaxy into a more spheroidal shape \citep{Emsellem2007}. A series of minor mergers can thus create a fast rotator, where the balance between dry and wet mergers determines the value of $\lambda_R$. Alternatively, as shown in hydrodynamical simulations,  major mergers can also result in a fast rotator, if the orbital and spin parameters are favourable \citep{Jesseit2009, Hoffman2010, Bois2011, Naab2014}.

The story is less clear for slow rotators. \citet{Robertson2006a} argue that slow-rotator progenitors must first have undergone gas-rich minor mergers to explain the tilt in the fundamental plane. To then remove angular momentum over time, various scenarios are possible. A series of dry minor mergers can decrease $\lambda_R$ as for the fast rotators. Alternatively, major mergers with discs of opposite spin are possible \citep{Puerari2001, Cappellari2007, Jesseit2009, Hoffman2010, Bois2011, Naab2014}. For the most massive slow rotators, which are found at the center of clusters \citep{Krajnovic2011}, a sequence of dry mergers is the most likely scenario. Galaxies that merge with the slow rotator will automatically have lower gas fractions due to environmental stripping of their gas.
 
In this context we will show in this letter that NGC 5626 is a peculiar case, exhibiting kinematic and photometric properties of both slow and fast rotators. This galaxy is indeed a regular rotator with an absolute K-band magnitude of $M_K = -25.5$, which already is as massive as fast rotators get. More strikingly, its velocity field shows a strong kinematic misalignment, which is highly unusual for fast rotators.

Throughout this work, we adopt a distance of 98.4 Mpc for the galaxy.

\section{Data \& Stellar Kinematics Extraction}

We have acquired MUSE \citep{Bacon2010} integral field spectroscopy for NGC 5626 during the instrument's Science Verification phase (run 60.A-9325(A)). A total integration time of 2 hours on source was split into four exposures, rotated by 90 degrees each time. The data were reduced using the standard pipeline, version 1.0. We refer to \citet{Viaene2017} for further details on the data acquisition and processing.

Before extracting the stellar and gas kinematics of NGC~5626 using the pPXF and GandALF programs of \citet{Cappellari2004} and \citet{Sarzi2006}, we proceeded to spatially bin the background-subtracted MUSE cube. First, we quantify the precision on the velocity and velocity dispersion as a function of the S/N of the spectra. For our science goals, a sufficient precision of 11 and 14 $\rm km\,s^{-1}$ (on the stellar velocity and velocity dispersion, respectively) is reached for a S/N of 30.
This target was then achieved using the Voronoi-binning procedure decribed in \citet{Cappellari2003}.

We then followed the approach of \citet{Sarzi2010} and \citet{Emsellem2014}, extracting nine high-S/N aperture spectra across different regions in NGC 5626 (from the center to the outermost regions probed by our MUSE observations) and obtained high-quality pPXF and GandALF fits to these spectra using the entire MILES stellar library \citep{Sanchez2006, Falcon2011}. In this way we also derive two sets of nine optimal stellar templates (one from the pPXF fit the other from the GandALF fit, differing essentially in their wavelength range). These two sets of optimal templates are then used to perform a fast and accurate fit to each Voronoi-bin. Throughout our analysis the pPXF and GandALF fits were done in the $5050 - 6000$ \AA\ and $4750 - 7500$ \AA\ wavelength regions, respectively. Fig.~\ref{fig:maps} shows the final pPXF maps for the stellar velocity and velocity dispersion, together with the `white-light' MUSE reconstructed image of NGC 5626. For the GandALF gas kinematics we refer the reader to Fig.~3 of \citet{Viaene2017}. High-order velocity moments and emission-line fluxes are also available.

\section{Analysis \& Results} 

\begin{figure}
    \subfloat{
    		\includegraphics[width=0.4\textwidth]{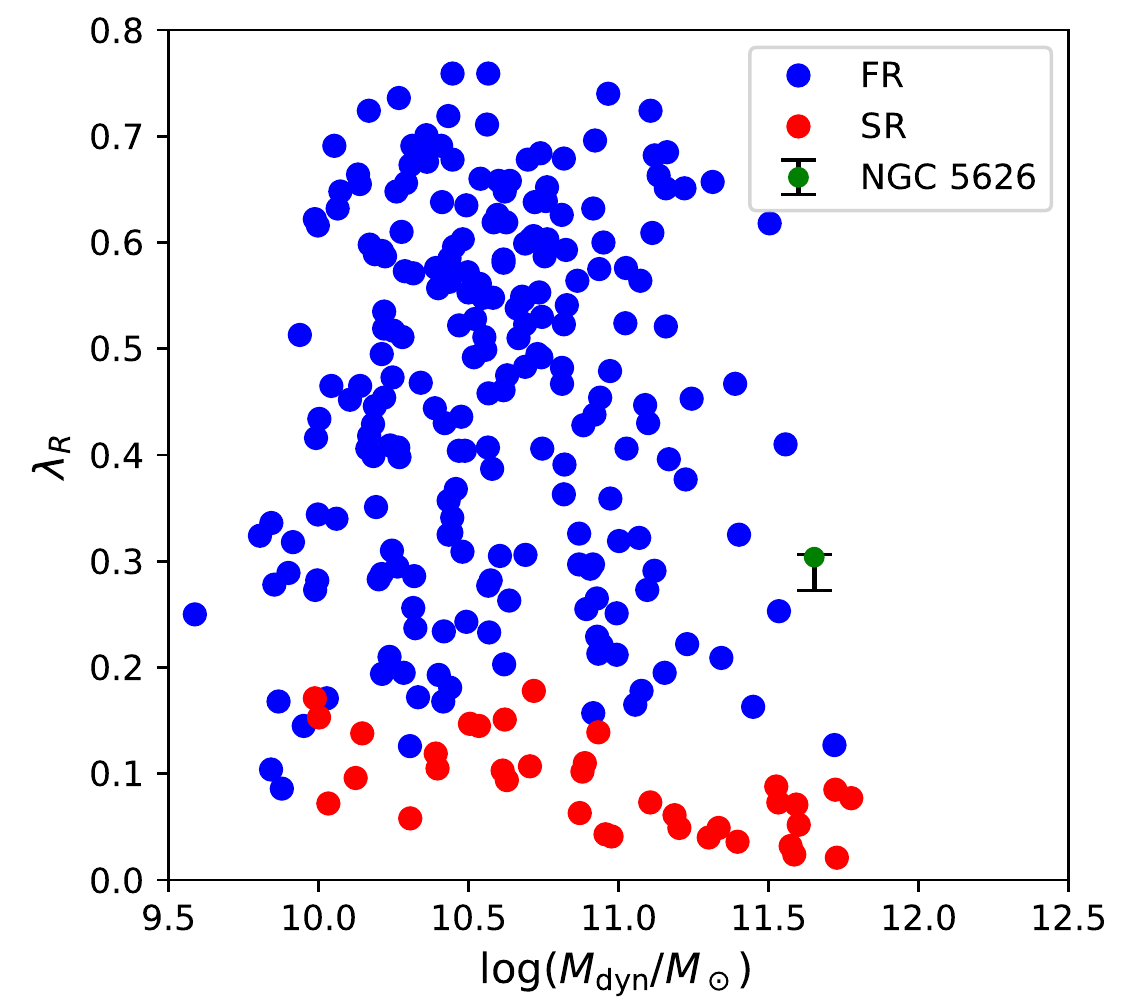} 
    }
    \hfill
    \vspace{-0.3cm}
    \subfloat{
    		\includegraphics[width=0.4\textwidth]{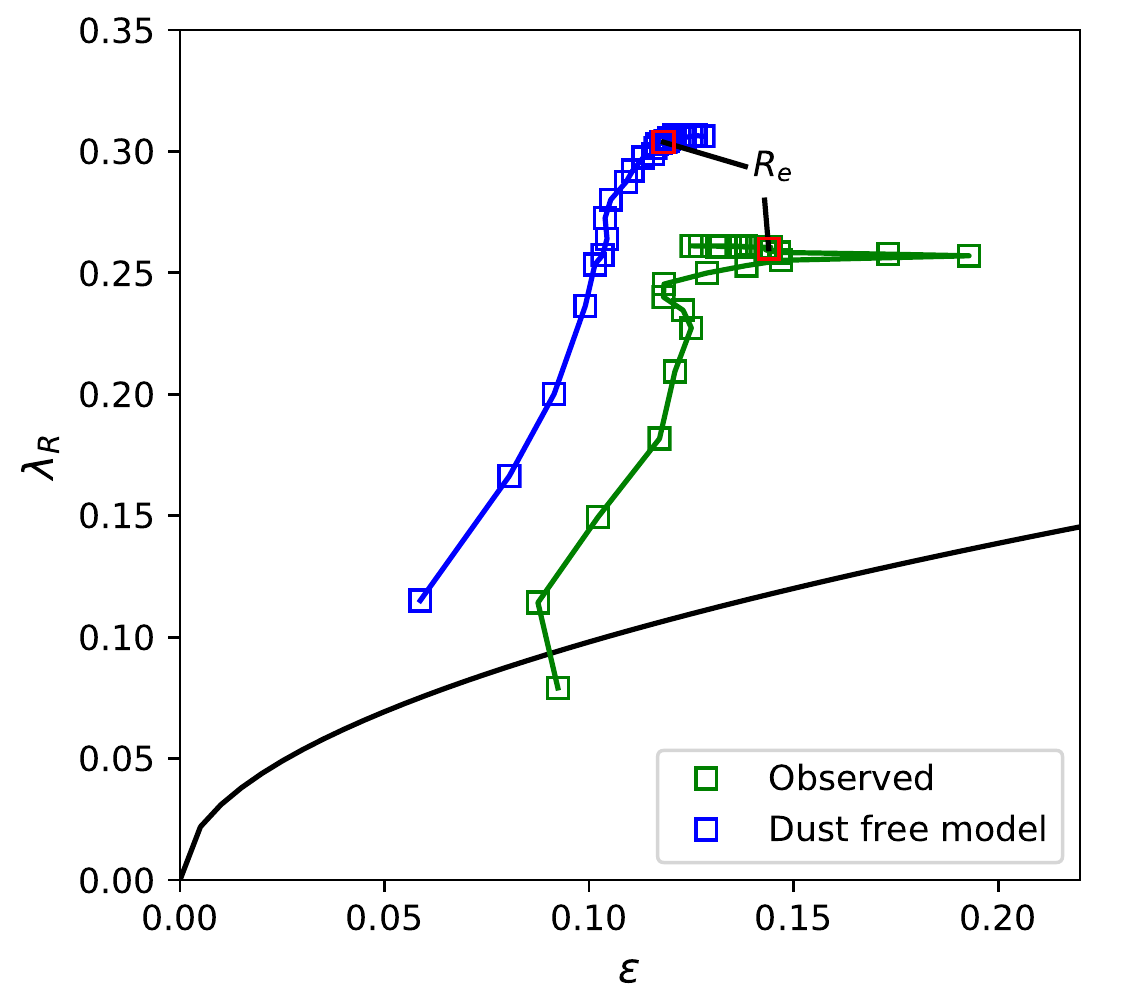} 
    }
\caption{Top: $\lambda_R$ vs. dynamical mass estimates. ATLAS$^\mathrm{3D}$ fast and slow rotators are coloured in blue and red, respectively. The dynamical mass for NGC 5626 was found to be $\log(M_\mathrm{vir}/M_\odot) = 11.7$, and plotted in green. The $\lambda_R$ value for NGC 5626 is the one from the dust-free model at $1R_e$. The error bar indicates the difference in $\lambda_R$ when exploring the $1-1.5R_e$ range, and changing between observed image and dust-free model.
Bottom: specific angular momentum $\lambda_R$ vs. apparent flattening $\epsilon$. Green points are measurements for NGC 5626 in radial bins of 1 arcsec (going roughly from bottom left to top right). The blue points show the measurements for the dust free model of NGC 5626 from \citet{Viaene2017}. The values for $1R_e$ are highlighted in red. The black line corresponds to the division line for fast and slow rotators as proposed by \citet{Emsellem2011}.}
    \label{fig:lambdaR_all}
\end{figure}

We first place NGC 5626 in the context of the ATLAS$^\mathrm{3D}$ sample. There are no morphological disturbances around the galaxy, so we can assume that the virial mass is a good approximation of the dynamical mass. As such, we can use the prescription of \citet{Cappellari2006}, where $M_\mathrm{vir} \approx 5.0 R_e \sigma_e^2/G$. We find $M_\mathrm{vir} \approx 5.4\times 10^{11}\, M_\odot$, which is at the high end of the ATLAS$^\mathrm{3D}$ mass distribution (Fig.~\ref{fig:lambdaR_all}). In fact, it lies above the threshold of $2\times10^{11} M_\odot$, set by \citet{Cappellari2016}, beyond which slow rotators dominate in number.

The MUSE field-of-view is too small to accurately determine the effective radius of the galaxy. We therefore chose to use the same $R_e$ formula as used for ATLAS$^\mathrm{3D}$ galaxies \citep[see eq. (6) in][]{Cappellari2011a}. This relies on the median $R_e=10.8^{\prime\prime}$ listed in the 2MASS-XSC \citep{Jarrett2000}, multiplied by $1.7$ to account for the 2MASS sensitivity limits. This brings the effective radius for NGC 5626 at $R_e=18.3$ arcsec.

To quantify the rotational state of NGC 5626, we follow the objective approach of \citet{Emsellem2011}. The ellipticity, $\epsilon$ was computed for each radial bin in the total flux image. Based on the ellipticity, we then determined the specific angular momentum, $\lambda_R$, in concentric elliptical regions out to $1.5R_e$. We plot all radial measurements in the $\lambda_R-\epsilon$ diagram in Fig.~\ref{fig:lambdaR_all}. 

The dust lane in NGC 5626 reduces the flux in the dust lane, and induces an apparent flattening of the isophotes, thus affecting the determination of  $\epsilon$ and $\lambda_R$. The green curve in the botom panel of Fig.~\ref{fig:lambdaR_all} indeed shows that $\lambda_R$ increases quickly with radius (as expected), but then the trend suddenly reverses and the ellipticity  decreases beyond $1R_e$. We therefore perform the same analysis on the dust free model of NGC 5626. This model was presented in \citet{Viaene2017} and relies on Multi-Gaussian Expansion fitting \citep{Cappellari2002} of the regions without dust extinction. The result is a shift towards lower $\epsilon$ as the dust lane no longer causes an apparent flattening of the stellar distribution. The evolution of $\lambda_R$ is also more natural, gradually rising with $\epsilon$ and galactocentric radius \citep[see e.g.][]{Emsellem2011}.  These authors also propose a division line to classify fast and slow rotators (black line in Fig.~\ref{fig:lambdaR_all}). At one effective radius, we obtain $\lambda_R = 0.31$ and $\epsilon = 0.12$, which unambiguously classifies NGC 5626 as a fast rotator.

The determination of $\lambda_R$ also allows us to place the galaxy in the  $\lambda_R-M_\mathrm{dyn}$ plane, together with the ATLAS$^\mathrm{3D}$ sample (Fig.~\ref{fig:lambdaR_all}, top panel). NGC 5626 again stands out as one of the most massive of the galaxies. Only one fast rotator is more massive (M60), but has a low $\lambda_R$ and is on the border of slow rotator classification.

\begin{figure*}
		\includegraphics[width=1.0\textwidth]{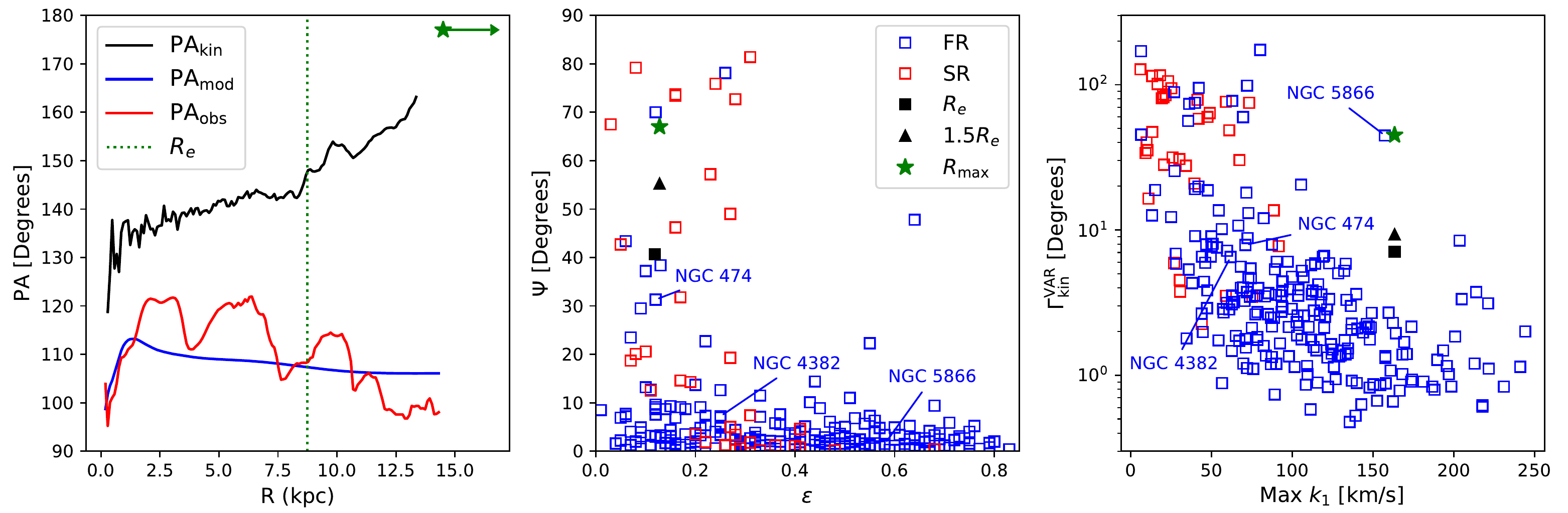}
    \caption{Left: Radial profiles of PA$_\mathrm{kin}$ as derived from the tilted ring fit, and photometric position angles from the observed (PA$_\mathrm{obs}$) and dust-free (PA$_\mathrm{mod}$) image. Middle: the kinematic misalignment $\Psi$ vs. ellipticity $\epsilon$. Right: variation of the kinematic PA ($\Gamma^\mathrm{VAR}_\mathrm{kin}$) vs. the maximum velocity $k_1$. The black square and triangle correspond to $1$ and $1.5R_e$ in NGC 5626. The green star is an estimate for the outermost radius ($R_\mathrm{max}$), based on the kinematic minor axis. ATLAS$^\mathrm{3D}$ fast and slow rotators are coloured in blue and red, respectively. Labelled galaxies are discussed in the text.}
    \label{fig:misalignment}
\end{figure*}

A striking outcome of our stellar-kinematics extraction is a high-quality map of the rotational velocity, which exhibits a strong twist towards the outer regions (Fig.~\ref{fig:maps}). We perform a tilted-ring model fit \citep[for details, see e.g.][]{Rogstad1974,Battaglia2006} to the velocity map to quantify the radial variation in PA$_\mathrm{kin}$. Fig.~\ref{fig:misalignment} (left panel) shows this radial profile, along with the ones for PA$_\mathrm{phot}$, derived using the standard \texttt{IRAF-ellipse} routine on the observed and dust-free white-light image. PA$_\mathrm{kin}$ increases linearly with radius within $1R_e$ and the rise accelerates beyond that. In contrast, PA$_\mathrm{phot}$ shows a slight decline. The radial profiles are limited to $1.5R_e$ in order to have the full ellipse in the FOV. However, along the kinematic minor axis (green colour in Fig.~\ref{fig:maps}, middle), we can visually trace the twist out to $\sim 1.8R_e = R_\mathrm{max}$. We estimate the position angle of the kinematic minor axis to be $\sim 90 \degree$ at this radius. Adding $90 \degree$ to obtain the PA of the kinematic major axis, this projection is shown in Fig.~\ref{fig:misalignment}.

The significant twist also leads to a large kinematic misalignment $\Psi$, which is defined as the difference between PA$_\mathrm{kin}$ and PA$_\mathrm{phot}$. This misalignment, which persists also in the central region, is shown in the context of ATLAS$^\mathrm{3D}$ in Fig.~\ref{fig:misalignment} (middle panel). The misalignments for these galaxies were computed using kinemetry \citep{Krajnovic2011}. For NGC 5626, a kinemetry fit provides good agreement with the tilted-ring method up to $1R_e$, but does not accurately capture the sharp PA$_\mathrm{kin}$ increase beyond that radius. Hence, we show only the tilted-ring results for NGC 5626. Among the fast rotators, only a few have a kinematic misalignment as large as the one we observe in NGC 5626, in particular beyond $R_e$. In fact, many of the fast rotators that exhibit a misalignment between the kinematic and photometric major axis are either barred or interacting. This makes NGC 5626 even more special as a fast rotator since it is neither barred nor currently interacting. 

To quantify kinematic variations such as twists, \citet{Krajnovic2011} defined $\Gamma^\mathrm{VAR}_\mathrm{kin}$, the standard deviation in the kinematic position angles, measured in radial bins up to $R_e$ (see their figure 6). We find $\Gamma^\mathrm{VAR}_\mathrm{kin} = 7.1-9.3 \degree$ between $1-1.5R_e$. On a plot of $\Gamma^\mathrm{VAR}_\mathrm{kin}$ vs. the maximum rotational velocity within those radii (max $k_1$), the galaxy immediately pops up as an outlier to the main trend (Fig.~\ref{fig:misalignment}, right panel). This is even more evident when assuming a variation of $\sim 45 \degree$, as the kinematic minor axis suggests. The only other galaxy near this location in the plot is NGC 5866, an edge-on Sa galaxy, though only the nucleus was covered in ATLAS$^\mathrm{3D}$. The map does not show an obvious twist as in NGC 5626 and it is likely that the $\Gamma^\mathrm{VAR}_\mathrm{kin}$ only reflects intrinsic uncertainty measurements.

There are only two ATLAS$^\mathrm{3D}$ galaxies identified as showing regular stellar rotation and smooth kinematic twist: NGC 474 and NGC 4382. Both features are weak compared to NGC 5626, and these systems are no outliers in Fig.~\ref{fig:misalignment}. Both galaxies contain shells and stellar streams in their outer halo, a clear morphological sign of a recent major interaction. This has not been observed in NGC 5626, though the presence of a dust lane hints at a minor merger \citep{Kaviraj2012}. We also looked for similar kinematic maps in the CALIFA sample of \citet{FalconBarroso2017} but did not find a twist as pronounced as the one in NGC 5626. Two galaxies with a kinematic twist were reported in the MASSIVE survey by \citet{Veale2017}: NGC 1129 and NGC 4874. However, the  coarse spatial resolution of their velocity maps make it hard to validate this. In any case, both galaxies are slow rotators and have $M_K < -26$, which is significantly more massive than NGC 5626.

Finally, we quantify the environmental density for NGC 5626 using the three metrics proposed in \citet{Cappellari2011}: $\rho_{10}$, the density of a sphere containing the 10 nearest neighbours; $\Sigma_{10}$, the surface density of a cilinder with height $-300 < \Delta v_\mathrm{hel} < 300\; \mathrm{km/s}$ and radius containing the 10 nearest neighbours in the cilindrical volume; $\Sigma_3$, as for $\Sigma_{10}$, but for the 3 nearest neighbours. We find $\rho_{10} = 0.28\; \mathrm{Mpc}^{-3}$, $\Sigma_{10} = 2.9\; \mathrm{Mpc}^{-2}$, and $\Sigma_3 = 8.9\; \mathrm{Mpc}^{-2}$. All three metrics indicate that NGC 5626 lies in an environment of intermediate density, at the edge of Virgo Cluster densities. \citet{Cappellari2011} argue that galaxies in this regime are likely to have undergone morphological segregation at the small group level. This can include dynamical processes such as cold accretion, close encounters or mergers, but also secular evolution \citep[see also][]{Kormendy2004, Debattista2006, Boselli2006}.

\section{Discussion \& conclusions} \label{sec:conclusions}

In short, NGC 5626 has properties of both slow and fast rotators. On the one hand, it is very massive and shows a clear kinematic twist. On the other hand, its velocity field is regular and specific angular momentum is relatively high.

Is NGC 5626 truly an exceptional case? The twist is most obvious beyond one effective radius, which means it could have been missed in the ATLAS$^\mathrm{3D}$ sample due to the limited field of view of the SAURON observations. In fact, for massive galaxies with $M_\mathrm{dyn} > 10^{11} M_\odot$ the ATLAS$^\mathrm{3D}$ observations extend typically out to $0.6 R_e$, whereas for NGC 5626 our MUSE data reach at least to $\sim 1.5R_e$. Outer twists in fast rotators are also not found in the CALIFA \citep{FalconBarroso2017} or MASSIVE \citep{Veale2017} kinematic maps, despite their larger field-of-view. However, MUSE provides a wider spectral range, and is mounted on a larger telescope than the aforementioned instruments, which allows to better sample the stellar kinematics in the fainter outskirts of galaxies. It is therefore unlikely that galaxies similar to NGC 5626 would be noticed in previous studies. As such, NGC 5626 may not be so unique as it seems today.

Another question is how this galaxy was formed and whether it had a slow or fast rotating progenitor. In fact, though large-scale jumps in the velocity field are often observed among slow-rotators, these correspond to either a) kinematically-decoupled cores and thus to an abrupt change in the kinematic position angle, or b) to two distinct peaks in the velocity dispersion map (which is not the case for NGC 5626) that in fact occur from the superposition of two counter-rotating stellar components in rather flat systems. Alternatively, it is likely that the smooth kinematic twist observed in NGC 5626 may result from an intrinsic triaxial shape for this galaxy \citep[e.g.][]{Arnold1994, Mathieu1999}.

In a series of merger simulations, \citet{Bois2011} are able to generate velocity twists for a few specific configurations. When merging a progenitor Sb galaxy with a smaller, gas-rich Sc galaxy (mass ratio of 2:1) in retrograde orbit, they obtain a large-scale kinematic twist only when decreasing the impact parameter to 35 kpc (a simulation they label `\textit{m21g33rdRm}'). In other words, a smooth twist occurs when the galaxies are positioned closer together at the start of their simulation, mimicking a more head-on collision. The merged galaxy also has a $\lambda_R$ of $\sim0.2$ at $R_e$, only slightly below to our measurement of $0.27$ for NGC 5626. Other merger configurations, of the same progenitors, yield instead either kinematically decoupled cores or more disturbed structure in the outer velocity field.  

This merging scenario suggests that a galaxy like NGC 5626 could have formed in a gas-rich merger, where the collision was relatively head-on, and could have transformed the galaxy's outer structure to a triaxial shape. Morphologically, the outskirts of NGC 5626 probed by our MUSE observations look rather undisturbed, suggesting that such a major merger happened at least more than $\sim 1$ Gyr ago, the dynamical timescale at 15 kpc from the center. Deeper images than those provided by 2MASS should be able to establish whether a possible merger happened further down in the past. The dust lane and ionised gas disc \citep[see][]{Viaene2017} in the inner part could then be the remnant of such a major event. However, it may also come from a more recent interaction \citep[see e.g.][]{Kaviraj2012}. The MUSE spectra indicate some star-formation activity in the dust lane. While a stellar population analysis in these regions could provide some constraint on the epoch of the accretion event that lead to the dust lane, tighter constraints on the assembly history of the outer parts of NGC 5626 (e.g. from deep imaging) would still be needed to distinguish between the above scenarios. 

Taken together, the possible merger history shows remarkable resemblance to the Milky Way - Andromeda collision scenario described by \citet{Cox2008}. They adopt the comparable masses and mass ratio as the `\textit{m21g33rdRm}' simulation from \citet{Bois2011} and the total baryonic mass in the simulation $(\log(M/M_\odot) = 11.1)$ lies only slightly below that of NGC 5626. Furthermore, both simulations start at the same infall speed, and are relatively head-on. Unfortunately, neither include dust particles which makes it unable to verify the existence of a settled dust disk, or its survival time. However, one can speculate that a merger between the Milky Way and Andromeda produces a galaxy of the same class as NGC 5626.

\section*{Acknowledgements}

Based on observations collected at the European Organisation for Astronomical Research in the Southern Hemisphere under ESO programme 60.A-9337(A). SV received support of the UGent BOF fund and the FWO mobility fund. 


\bibliographystyle{mnras}
\bibliography{references} 

\begin{thebibliography}{}
\makeatletter
\relax
\def\mn@urlcharsother{\let\do\@makeother \do\$\do\&\do\#\do\^\do\_\do\%\do\~}
\def\mn@doi{\begingroup\mn@urlcharsother \@ifnextchar [ {\mn@doi@}
  {\mn@doi@[]}}
\def\mn@doi@[#1]#2{\def\@tempa{#1}\ifx\@tempa\@empty \href
  {http://dx.doi.org/#2} {doi:#2}\else \href {http://dx.doi.org/#2} {#1}\fi
  \endgroup}
\def\mn@eprint#1#2{\mn@eprint@#1:#2::\@nil}
\def\mn@eprint@arXiv#1{\href {http://arxiv.org/abs/#1} {{\tt arXiv:#1}}}
\def\mn@eprint@dblp#1{\href {http://dblp.uni-trier.de/rec/bibtex/#1.xml}
  {dblp:#1}}
\def\mn@eprint@#1:#2:#3:#4\@nil{\def\@tempa {#1}\def\@tempb {#2}\def\@tempc
  {#3}\ifx \@tempc \@empty \let \@tempc \@tempb \let \@tempb \@tempa \fi \ifx
  \@tempb \@empty \def\@tempb {arXiv}\fi \@ifundefined
  {mn@eprint@\@tempb}{\@tempb:\@tempc}{\expandafter \expandafter \csname
  mn@eprint@\@tempb\endcsname \expandafter{\@tempc}}}

\bibitem[\protect\citeauthoryear{{Arnold}, {de Zeeuw}  \& {Hunter}}{{Arnold}
  et~al.}{1994}]{Arnold1994}
{Arnold} R.,  {de Zeeuw} P.~T.,   {Hunter} C.,  1994, \mn@doi [\mnras]
  {10.1093/mnras/271.4.924}, \href
  {http://adsabs.harvard.edu/abs/1994MNRAS.271..924A} {271, 924}

\bibitem[\protect\citeauthoryear{{Bacon} et~al.,}{{Bacon}
  et~al.}{2010}]{Bacon2010}
{Bacon} R.,  et~al., 2010. p. 773508, \mn@doi{10.1117/12.856027}

\bibitem[\protect\citeauthoryear{{Battaglia}, {Fraternali}, {Oosterloo}  \&
  {Sancisi}}{{Battaglia} et~al.}{2006}]{Battaglia2006}
{Battaglia} G.,  {Fraternali} F.,  {Oosterloo} T.,   {Sancisi} R.,  2006,
  \mn@doi [\aap] {10.1051/0004-6361:20053210}, \href
  {http://adsabs.harvard.edu/abs/2006A%26A...447...49B} {447, 49}

\bibitem[\protect\citeauthoryear{{Bois} et~al.,}{{Bois}
  et~al.}{2011}]{Bois2011}
{Bois} M.,  et~al., 2011, \mn@doi [\mnras] {10.1111/j.1365-2966.2011.19113.x},
  \href {http://adsabs.harvard.edu/abs/2011MNRAS.416.1654B} {416, 1654}

\bibitem[\protect\citeauthoryear{{Boselli} \& {Gavazzi}}{{Boselli} \&
  {Gavazzi}}{2006}]{Boselli2006}
{Boselli} A.,  {Gavazzi} G.,  2006, \mn@doi [\pasp] {10.1086/500691}, \href
  {http://adsabs.harvard.edu/abs/2006PASP..118..517B} {118, 517}

\bibitem[\protect\citeauthoryear{{Cappellari}}{{Cappellari}}{2002}]{Cappellari2002}
{Cappellari} M.,  2002, \mn@doi [\mnras] {10.1046/j.1365-8711.2002.05412.x},
  \href {http://adsabs.harvard.edu/abs/2002MNRAS.333..400C} {333, 400}

\bibitem[\protect\citeauthoryear{{Cappellari}}{{Cappellari}}{2016}]{Cappellari2016}
{Cappellari} M.,  2016, \mn@doi [\araa] {10.1146/annurev-astro-082214-122432},
  \href {http://adsabs.harvard.edu/abs/2016ARA%26A..54..597C} {54, 597}

\bibitem[\protect\citeauthoryear{{Cappellari} \& {Copin}}{{Cappellari} \&
  {Copin}}{2003}]{Cappellari2003}
{Cappellari} M.,  {Copin} Y.,  2003, \mn@doi [\mnras]
  {10.1046/j.1365-8711.2003.06541.x}, \href
  {http://adsabs.harvard.edu/abs/2003MNRAS.342..345C} {342, 345}

\bibitem[\protect\citeauthoryear{{Cappellari} \& {Emsellem}}{{Cappellari} \&
  {Emsellem}}{2004}]{Cappellari2004}
{Cappellari} M.,  {Emsellem} E.,  2004, \mn@doi [\pasp] {10.1086/381875}, \href
  {http://adsabs.harvard.edu/abs/2004PASP..116..138C} {116, 138}

\bibitem[\protect\citeauthoryear{{Cappellari} et~al.,}{{Cappellari}
  et~al.}{2006}]{Cappellari2006}
{Cappellari} M.,  et~al., 2006, \mn@doi [\mnras]
  {10.1111/j.1365-2966.2005.09981.x}, \href
  {http://adsabs.harvard.edu/abs/2006MNRAS.366.1126C} {366, 1126}

\bibitem[\protect\citeauthoryear{{Cappellari} et~al.,}{{Cappellari}
  et~al.}{2007}]{Cappellari2007}
{Cappellari} M.,  et~al., 2007, \mn@doi [\mnras]
  {10.1111/j.1365-2966.2007.11963.x}, \href
  {http://adsabs.harvard.edu/abs/2007MNRAS.379..418C} {379, 418}

\bibitem[\protect\citeauthoryear{{Cappellari} et~al.,}{{Cappellari}
  et~al.}{2011a}]{Cappellari2011a}
{Cappellari} M.,  et~al., 2011a, \mn@doi [\mnras]
  {10.1111/j.1365-2966.2010.18174.x}, \href
  {http://adsabs.harvard.edu/abs/2011MNRAS.413..813C} {413, 813}

\bibitem[\protect\citeauthoryear{{Cappellari} et~al.,}{{Cappellari}
  et~al.}{2011b}]{Cappellari2011}
{Cappellari} M.,  et~al., 2011b, \mn@doi [\mnras]
  {10.1111/j.1365-2966.2011.18600.x}, \href
  {http://adsabs.harvard.edu/abs/2011MNRAS.416.1680C} {416, 1680}

\bibitem[\protect\citeauthoryear{{Cox} \& {Loeb}}{{Cox} \&
  {Loeb}}{2008}]{Cox2008}
{Cox} T.~J.,  {Loeb} A.,  2008, \mn@doi [\mnras]
  {10.1111/j.1365-2966.2008.13048.x}, \href
  {http://adsabs.harvard.edu/abs/2008MNRAS.386..461C} {386, 461}

\bibitem[\protect\citeauthoryear{{Debattista}, {Mayer}, {Carollo}, {Moore},
  {Wadsley}  \& {Quinn}}{{Debattista} et~al.}{2006}]{Debattista2006}
{Debattista} V.~P.,  {Mayer} L.,  {Carollo} C.~M.,  {Moore} B.,  {Wadsley} J.,
   {Quinn} T.,  2006, \mn@doi [\apj] {10.1086/504147}, \href
  {http://adsabs.harvard.edu/abs/2006ApJ...645..209D} {645, 209}

\bibitem[\protect\citeauthoryear{{Emsellem} et~al.,}{{Emsellem}
  et~al.}{2007}]{Emsellem2007}
{Emsellem} E.,  et~al., 2007, \mn@doi [\mnras]
  {10.1111/j.1365-2966.2007.11752.x}, \href
  {http://adsabs.harvard.edu/abs/2007MNRAS.379..401E} {379, 401}

\bibitem[\protect\citeauthoryear{{Emsellem} et~al.,}{{Emsellem}
  et~al.}{2011}]{Emsellem2011}
{Emsellem} E.,  et~al., 2011, \mn@doi [\mnras]
  {10.1111/j.1365-2966.2011.18496.x}, \href
  {http://adsabs.harvard.edu/abs/2011MNRAS.414..888E} {414, 888}

\bibitem[\protect\citeauthoryear{{Emsellem}, {Krajnovi{\'c}}  \&
  {Sarzi}}{{Emsellem} et~al.}{2014}]{Emsellem2014}
{Emsellem} E.,  {Krajnovi{\'c}} D.,   {Sarzi} M.,  2014, \mn@doi [\mnras]
  {10.1093/mnrasl/slu140}, \href
  {http://adsabs.harvard.edu/abs/2014MNRAS.445L..79E} {445, L79}

\bibitem[\protect\citeauthoryear{{Falc{\'o}n-Barroso}
  et~al.,}{{Falc{\'o}n-Barroso} et~al.}{2011}]{Falcon2011}
{Falc{\'o}n-Barroso} J.,  et~al., 2011, \mn@doi [\aap]
  {10.1051/0004-6361/201116842}, \href
  {http://adsabs.harvard.edu/abs/2011A%26A...532A..95F} {532, A95}

\bibitem[\protect\citeauthoryear{{Falc{\'o}n-Barroso}
  et~al.,}{{Falc{\'o}n-Barroso} et~al.}{2017}]{FalconBarroso2017}
{Falc{\'o}n-Barroso} J.,  et~al., 2017, \mn@doi [\aap]
  {10.1051/0004-6361/201628625}, \href
  {http://adsabs.harvard.edu/abs/2017A%26A...597A..48F} {597, A48}

\bibitem[\protect\citeauthoryear{{Hoffman}, {Cox}, {Dutta}  \&
  {Hernquist}}{{Hoffman} et~al.}{2010}]{Hoffman2010}
{Hoffman} L.,  {Cox} T.~J.,  {Dutta} S.,   {Hernquist} L.,  2010, \mn@doi
  [\apj] {10.1088/0004-637X/723/1/818}, \href
  {http://adsabs.harvard.edu/abs/2010ApJ...723..818H} {723, 818}

\bibitem[\protect\citeauthoryear{{Hubble}}{{Hubble}}{1936}]{Hubble1936}
{Hubble} E.~P.,  1936, {Realm of the Nebulae}

\bibitem[\protect\citeauthoryear{{Jarrett}, {Chester}, {Cutri}, {Schneider},
  {Skrutskie}  \& {Huchra}}{{Jarrett} et~al.}{2000}]{Jarrett2000}
{Jarrett} T.~H.,  {Chester} T.,  {Cutri} R.,  {Schneider} S.,  {Skrutskie} M.,
   {Huchra} J.~P.,  2000, \mn@doi [\aj] {10.1086/301330}, \href
  {http://adsabs.harvard.edu/abs/2000AJ....119.2498J} {119, 2498}

\bibitem[\protect\citeauthoryear{{Jesseit}, {Cappellari}, {Naab}, {Emsellem}
  \& {Burkert}}{{Jesseit} et~al.}{2009}]{Jesseit2009}
{Jesseit} R.,  {Cappellari} M.,  {Naab} T.,  {Emsellem} E.,   {Burkert} A.,
  2009, \mn@doi [\mnras] {10.1111/j.1365-2966.2009.14984.x}, \href
  {http://adsabs.harvard.edu/abs/2009MNRAS.397.1202J} {397, 1202}

\bibitem[\protect\citeauthoryear{{Kaviraj} et~al.,}{{Kaviraj}
  et~al.}{2012}]{Kaviraj2012}
{Kaviraj} S.,  et~al., 2012, \mn@doi [\mnras]
  {10.1111/j.1365-2966.2012.20957.x}, \href
  {http://adsabs.harvard.edu/abs/2012MNRAS.423...49K} {423, 49}

\bibitem[\protect\citeauthoryear{{Kormendy} \& {Bender}}{{Kormendy} \&
  {Bender}}{1996}]{Kormendy1996}
{Kormendy} J.,  {Bender} R.,  1996, \mn@doi [\apjl] {10.1086/310095}, \href
  {http://adsabs.harvard.edu/abs/1996ApJ...464L.119K} {464, L119}

\bibitem[\protect\citeauthoryear{{Kormendy} \& {Kennicutt}}{{Kormendy} \&
  {Kennicutt}}{2004}]{Kormendy2004}
{Kormendy} J.,  {Kennicutt} Jr. R.~C.,  2004, \mn@doi [\araa]
  {10.1146/annurev.astro.42.053102.134024}, \href
  {http://adsabs.harvard.edu/abs/2004ARA%26A..42..603K} {42, 603}

\bibitem[\protect\citeauthoryear{{Krajnovi{\'c}} et~al.,}{{Krajnovi{\'c}}
  et~al.}{2011}]{Krajnovic2011}
{Krajnovi{\'c}} D.,  et~al., 2011, \mn@doi [\mnras]
  {10.1111/j.1365-2966.2011.18560.x}, \href
  {http://adsabs.harvard.edu/abs/2011MNRAS.414.2923K} {414, 2923}

\bibitem[\protect\citeauthoryear{{Mathieu} \& {Dejonghe}}{{Mathieu} \&
  {Dejonghe}}{1999}]{Mathieu1999}
{Mathieu} A.,  {Dejonghe} H.,  1999, \mn@doi [\mnras]
  {10.1046/j.1365-8711.1999.02199.x}, \href
  {http://adsabs.harvard.edu/abs/1999MNRAS.303..455M} {303, 455}

\bibitem[\protect\citeauthoryear{{Naab} et~al.,}{{Naab}
  et~al.}{2014}]{Naab2014}
{Naab} T.,  et~al., 2014, \mn@doi [\mnras] {10.1093/mnras/stt1919}, \href
  {http://adsabs.harvard.edu/abs/2014MNRAS.444.3357N} {444, 3357}

\bibitem[\protect\citeauthoryear{{Puerari} \& {Pfenniger}}{{Puerari} \&
  {Pfenniger}}{2001}]{Puerari2001}
{Puerari} I.,  {Pfenniger} D.,  2001, \mn@doi [\apss]
  {10.1023/A:1017581325673}, \href
  {http://adsabs.harvard.edu/abs/2001Ap%26SS.276..909P} {276, 909}

\bibitem[\protect\citeauthoryear{{Robertson}, {Cox}, {Hernquist}, {Franx},
  {Hopkins}, {Martini}  \& {Springel}}{{Robertson}
  et~al.}{2006}]{Robertson2006a}
{Robertson} B.,  {Cox} T.~J.,  {Hernquist} L.,  {Franx} M.,  {Hopkins} P.~F.,
  {Martini} P.,   {Springel} V.,  2006, \mn@doi [\apj] {10.1086/500360}, \href
  {http://adsabs.harvard.edu/abs/2006ApJ...641...21R} {641, 21}

\bibitem[\protect\citeauthoryear{{Rogstad}, {Lockhart}  \& {Wright}}{{Rogstad}
  et~al.}{1974}]{Rogstad1974}
{Rogstad} D.~H.,  {Lockhart} I.~A.,   {Wright} M.~C.~H.,  1974, \mn@doi [\apj]
  {10.1086/153164}, \href {http://adsabs.harvard.edu/abs/1974ApJ...193..309R}
  {193, 309}

\bibitem[\protect\citeauthoryear{{S{\'a}nchez-Bl{\'a}zquez}
  et~al.,}{{S{\'a}nchez-Bl{\'a}zquez} et~al.}{2006}]{Sanchez2006}
{S{\'a}nchez-Bl{\'a}zquez} P.,  et~al., 2006, \mn@doi [\mnras]
  {10.1111/j.1365-2966.2006.10699.x}, \href
  {http://adsabs.harvard.edu/abs/2006MNRAS.371..703S} {371, 703}

\bibitem[\protect\citeauthoryear{{Sarzi} et~al.,}{{Sarzi}
  et~al.}{2006}]{Sarzi2006}
{Sarzi} M.,  et~al., 2006, \mn@doi [\mnras] {10.1111/j.1365-2966.2005.09839.x},
  \href {http://adsabs.harvard.edu/abs/2006MNRAS.366.1151S} {366, 1151}

\bibitem[\protect\citeauthoryear{{Sarzi} et~al.,}{{Sarzi}
  et~al.}{2010}]{Sarzi2010}
{Sarzi} M.,  et~al., 2010, \mn@doi [\mnras] {10.1111/j.1365-2966.2009.16039.x},
  \href {http://adsabs.harvard.edu/abs/2010MNRAS.402.2187S} {402, 2187}

\bibitem[\protect\citeauthoryear{{Veale} et~al.,}{{Veale}
  et~al.}{2017}]{Veale2017}
{Veale} M.,  et~al., 2017, \mn@doi [\mnras] {10.1093/mnras/stw2330}, \href
  {http://adsabs.harvard.edu/abs/2017MNRAS.464..356V} {464, 356}

\bibitem[\protect\citeauthoryear{{Viaene}, {Sarzi}, {Baes}, {Fritz}  \&
  {Puerari}}{{Viaene} et~al.}{2017}]{Viaene2017}
{Viaene} S.,  {Sarzi} M.,  {Baes} M.,  {Fritz} J.,   {Puerari} I.,  2017,
  preprint, \href {http://adsabs.harvard.edu/abs/2017arXiv170706237V} {}
  (\mn@eprint {arXiv} {1707.06237})

\bibitem[\protect\citeauthoryear{{Weijmans} et~al.,}{{Weijmans}
  et~al.}{2014}]{Weijmans2014}
{Weijmans} A.-M.,  et~al., 2014, \mn@doi [\mnras] {10.1093/mnras/stu1603},
  \href {http://adsabs.harvard.edu/abs/2014MNRAS.444.3340W} {444, 3340}

\makeatother
\end{thebibliography}




\bsp	
\label{lastpage}
\end{document}